\documentclass[twocolumn]{aa}
\usepackage{graphicx}
\usepackage{natbib}
\usepackage{txfonts}
\begin{document}
%
\def\ltsima{$\; \buildrel < \over \sim\;$}
\def\ltsim{\lower.5ex\hbox{\ltsima}}
\def\gtsima{$\; \buildrel > \over\sim \;$}
\def\gtsim{\lower.5ex\hbox{\gtsima}}
\def\ms{$M_{\odot}$ }
\def\msp{$M_{\odot}$}
\title{Implications of a non-universal IMF from C, N, and O abundances in very metal-poor Galactic stars and damped Ly$\alpha$ absorbers}
\titlerunning{Implications of a non-universal IMF from C, N, and O abundances}
\authorrunning{Tsujimoto \& Bekki}

\author{ T. Tsujimoto\inst{1} \and K. Bekki\inst{2}}
\offprints{ T. Tsujimoto}
 \institute{\inst{1}National Astronomical Observatory of Japan, Mitaka,
                  Tokyo 181-8588, Japan (\email{taku.tsujimoto@nao.ac.jp}) \\
                  \inst{2}ICRAR, M468, The University of Western Australia, 35 Stirling Highway, Crawley
                  Western Australia 6009, Austraiia}

\date{Received 26 November 2010 / Accepted 14 February 2011}    

\abstract{Recently revealed C, N, and O abundances in the most metal-poor damped Ly$\alpha$ (DLA) absorbers are compared with those of extremely metal-poor stars in the Galactic halo, as well as extragalactic H II regions, to decipher nucleosynthesis and chemical enrichment in the early Universe. These comparisons surprisingly identify a relatively high C/O ratio and a low N/O ratio in DLA systems, which is hard to explain theoretically. We propose that if these features are confirmed by future studies, this effect occurs because the initial mass function in metal-poor DLA systems has a cut-off at the upper mass end at around 20--25 \msp, thus lacks the massive stars that provide the nucleosynthesis products leading to the low C/O and high N/O ratios. This finding is a reasonable explanation of the nature of DLA systems in which a sufficient amount of cold H I gas remains intact because of the suppression of ionization by massive stars. In addition, our claim strongly supports a high production rate of N in very massive stars, which might be acceptable in light of the recent nucleosynthesis calculations with fast rotation models. The updates of both abundance data and nucleosynthesis results will strengthen our novel proposition that the C/O and N/O abundances are a powerful tool for inferring the form of the initial mass function.
}

\maketitle
%
 
\section{Introduction}
Precise measurements of elemental abundances in the early Universe enable us to have direct access to the chemical enrichment process by studying nucleosynthesis in massive stars. Now we have two different kinds of available targets, i.e, extremely metal-poor (EMP) stars in the Galactic halo \citep[e.g.,][]{Cayrel_04} and high-z damped Ly$\alpha$ systems (DLAs) (e.g., Pettini et al. 2002, 2008). Their metallicity ranges are approximately -4$<$[Fe/H]$<$-2.5 and -3$<$[Fe/H]$<$-2, respectively. Several theoretical works claim that the abundances of EMP stars retain the nucleosynthesis yields of individual type II supernovae (SNe II) \citep{Audouze_95, Shigeyama_98, Nakamura_99, Umeda_02}. In contrast, DLAs represent the abundances of the interstellar medium (ISM) integrated along the line of sight, which is equivalent to an average of all the mass of SNe II weighted with an initial mass function (IMF). 

\citet{Pettini_02} measured the N and O abundances of metal-poor DLAs \citep[see also][]{Centurion_03}. It is well-known that the N/O ratio of extragalactic H II regions exhibits a plateau at  12+$\log$ O/H \ltsim 8.0 [Note that the solar O abundance is 8.69 (Asplund et al.~2009)], which is considered to represent the primary production level of N. Their finding is, however,  that some DLAs in their samples have a lower N/O ratio than the plateau level. This was firmly confirmed by \citet{Pettini_08} with more additional samples (see left panel of Fig.~1). It is possible to infer from this that the observed low N/O ratio is attributable to the time delay of N ejection from asymptotic giant branch (AGB) stars with respect to O production by SNe II, provided that the primary N is mainly produced in AGB stars and star formation proceeds at a lower rate in DLAs than in extragalactic HII regions \citep{Calura_03, Molla_06, Henry_07}. 

However, these predictions are incompatible with the observed N/O ratios of EMP stars. \citet{Spite_05} determined the N abundance of red giants, which is not influenced by any mixing inside evolved stars. They utilized the Li abundance measured in individual stars to diagnose the mixing level. Their result is shown by squares in the left panel of Fig. 1. We can see that an average N/O ratio of stars is broadly consistent with the plateau value. That implies that (i) massive stars synthesize a large amount of primary N, and (ii) the assembly of primary N from massive stars produces the plateau exhibited by extragalactic HII regions.

This high N abundance observed in EMP stars has inspired renewed interest in the nucleosynthesis of N in massive stars, which has led to the discovery of a new channel that enables high N production in rapidly rotating massive stars \citep{Chiappini_06, Hirschi_07, Ekstrom_08}. Leaving more progress on this issue to future studies, the question is then how we interpret the low N/O ratio in DLAs if enough N is synthesized in massive stars to reach the level equivalent to the N/O ratios of EMP stars. 

  \begin{figure*}
   \centering
  \includegraphics[angle=0,width=0.9\textwidth]{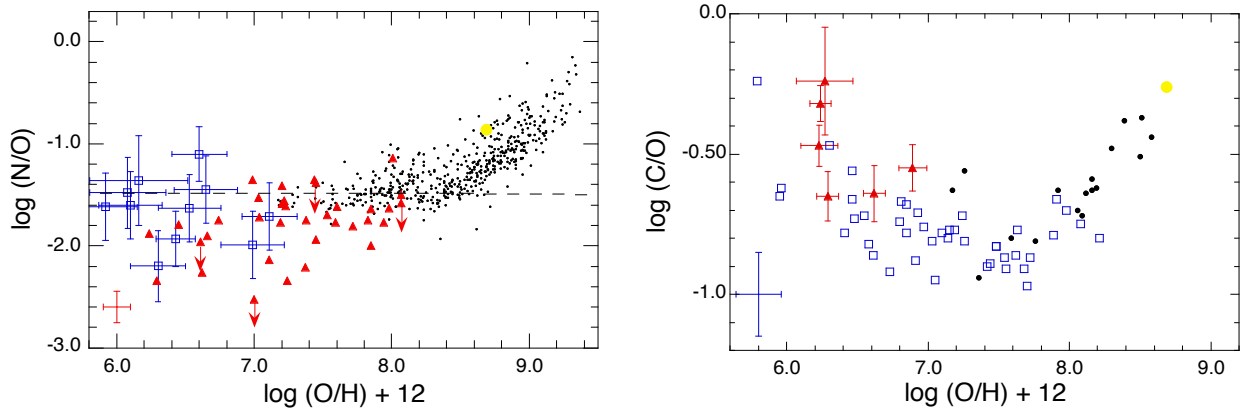}
       \caption{{\it Left panel}: Correlation of N/O with O/H in DLA systems \citep[triangles:][]{Pettini_08}, Galactic halo stars \citep[squares:][]{Spite_05}, and extragalactic H II regions (small dots). The data of H II regions are from the sources assembled by \citet{Pettini_08}. The error bar in the bottom corner is for DLAs. The dashed line is the approximate plateau level seen in extragalactic H II regions. The large yellow dot corresponds to the solar abundances \citep{Asplund_09}. {\it Right panel}: Correlation of C/O with O/H. The data of H II regions are from Garnett et al. (1995, 1997, 1999) and Kobulnicky et al. (1997). The Galactic values are from \citet{Fabbian_09a}. The error bar in the bottom left-hand corner gives an indication of the typical uncertainty for halo stars.  
}
  \label{}
   \end{figure*}

A clue to tackling this enigma can be found in the C abundance of DLAs, which is measured together with N and O abundances \citep{Pettini_08}. As shown in the right panel of Fig. 1, the C/O ratios in DLAs (triangles) are somewhat higher than those of EMP stars (squares), which is the opposite of the trend found for the N/O ratio. The origin of C/O feature is indeed far easier to access than that of N/O because we have a much clearer understanding of the nucleosynthesis of C than N in massive stars \citep{Shigeyama_98}. 

In this paper, we propose a unified scheme that explains C, N, and O abundances in both DLAs and EMP stars, as well as in extragalactic HII regions. Our major claim is that if the IMF lacks very massive stars beyond $\sim$ 20-25 \ms in metal-poor DLAs for some physical reasons that  will be discussed in detail, both the observed high C/O and low N/O ratios are consistently explained. 
A theoretical scheme that does not invoke changes in the IMF based on the idea that DLAs have a different speed of chemical enrichment does not sufficiently explain their abundances, since this view predicts the same direction of change in both the C/O and N/O ratios against those of EMP stars. A model in which the major contribution of Population (Pop) III stars to chemical abundances is considered \citep{Akerman_04} is not likely either, because the N /O ratios of EMP stars seem to reflect no clear difference in the production of N between Pop.~II and Pop.~III stars. In addition, a metallicity-dependent IMF, as well as metallicity-dependent yields among C, N, and O,  may have to be considered as the origin of abundance trends in the Galactic halo but not as the mechanism of the contrast between DLAs and EMP stars.

The proposed IMF is supported by an observational result that suggests that the observed flux ratio of H$\alpha$ to the far-ultraviolet in low surface-brightness (LSB) galaxies favors such a truncated IMF \citep{Meurer_09}, though there seems to be a wide range of other explanations for the observed trend (Lee et al., 2009, 2011a). In contrast to these truncated IMFs, there is observational evidence of a top-heavy IMF in massive galaxies at high redshifts \citep[e.g.,][]{Dokkum_08}. Accordingly, we insist on a non-universal IMF that varies among different galactic objects.

\section{How different are DLAs and stars?}

The argument presented in this paper is based on the observationally implied contrast between the relative abundances of C, N, and O in Galactic halo stars and metal-poor DLAs. However,  as shown in Fig. 1, mainly because of the large errors in each set of data, the extents of their differences seem insignificant and debatable. In this section, we do the statistical tests and also discuss the uncertainties involved in the abundance determination that may yield the systematic errors.

\subsection{C/O}

We assess the extent of the differences in C/O between the two data groups, including the effect of the random error in the observed abundances of C/O and O/H of individual stars and DLAs. We compare all DLAs with the stars residing in $\log$ (O/H)+12$<$7.0 to ensure that both of the metallicity ranges broadly coincide. Since the C/O ratio of stars has a clear trend against O/H, we perform the least squares fit and see the C/O value at $\log$ (O/H)+12$=$6.5. We take this O/H value as the zero-point of a straight line. The same procedure is performed for DLAs and also for N/O in Sect. 2.2. As a result, we obtain $\log$ (C/O)=-0.68$\pm$0.04 for stars and  -0.51$\pm$0.04 for DLAs.

  \begin{figure*}
   \centering
  \includegraphics[angle=0,width=0.8\textwidth]{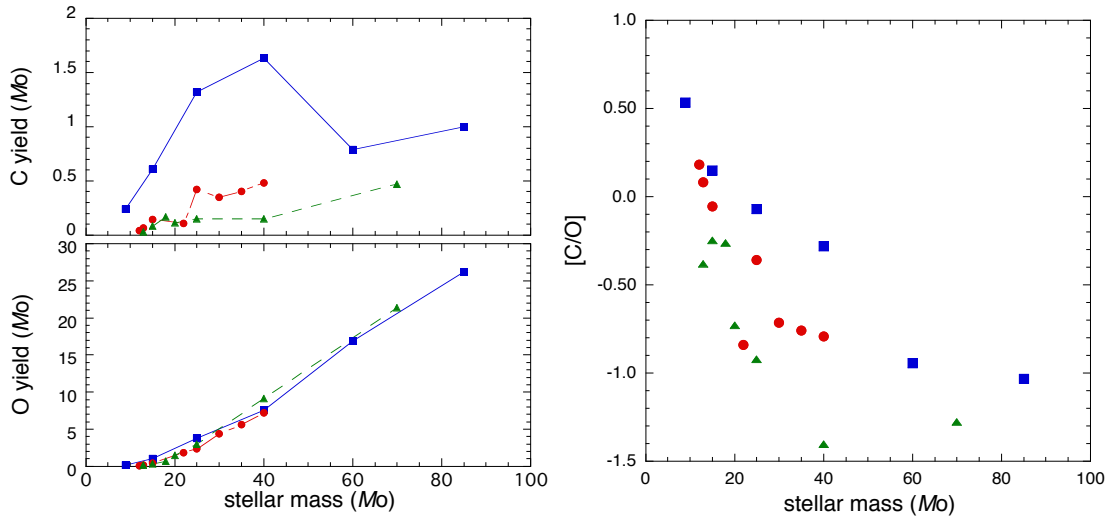}
       \caption{{\it Left panels}: C and O yields of massive stars as a function of the progenitor mass predicted by three theoretical models (circles: Woosley \& Weaver 1995; triangles: Nomoto et al. 1997; squares: Ekstr\"{o}m et al. 2008). {\it Right panel}: Theoretical nucleosynthesis [C/O] ratio predicted by the three models as a function of the progenitor mass.  
}
  \label{}
   \end{figure*}

In addition to random errors, there are systematic uncertainties because the stellar and DLA abundances are derived by different methods. Most notably, the efficiency of inelastic H collisions, whose cross-section remains essentially unknown, influences the values of stellar C/O and O/H. Here we adopt the case of a high collision efficiency that is adopted as a standard case in \citet{Fabbian_09a}. If these collisions are unimportant, stellar and DLA abundances are in broad agreement \citep{Fabbian_09a}. However, this results in an unusual trend of [O/Fe] against [Fe/H]: the [O/Fe] ratio of halo stars decreases towards lower  [Fe/H] for [Fe/H]$<$-2.0 \citep{Fabbian_09b}, which is inconsistent with the elemental ratios of other $\alpha$-elements to Fe. Moreover, a finding of a very high C abundance in a DLA with [Fe/H]=-3 \citep{Cooke_10} is likely to agree with  the contrast of C/O between DLAs and Galactic stars.

\subsection{N/O}

We compare all EMP stars with the DLAs residing in $\log$ (O/H)+12$<$7.2 for the same reason as in the C/O case. At $\log$ (O/H)+12$=$6.5, $\log$ (N/O)=-1.64$\pm$0.1 for stars and -2.1$\pm$0.07 for DLAs are deduced. Here for two DLAs, we assume their mean values to be their upper limits. The difference between stars and DLAs becomes smaller when we exclude the four stars at $\log$ (O/H)+12$<$6.2, where there is no data for DLAs. This stellar ensemble then gives $\log$ (C/O)=-1.82$\pm$0.23.

For the systematic errors, we note that the stellar abundance of N is derived from the ultraviolet NH band, whose physical parameters are not yet well established \citep{Spite_05}. Thus, the possibility  of the stellar N/O ratios being systematically somewhat smaller than those in Fig. 1 cannot be ruled out. The N/O ratio measured in unevolved dwarf stars \citep{Israelian_04} includes three stars in our metallicity range, i.e., $\log$ (O/H)+12$<$7.2. Their average N/O value is  $\log$ (N/O)=-1.49, which seems broadly consistent with the analysis of \citet{Spite_05}. Moreover, if we add three samples to Spite's data in the range of 6.2$<$$\log$ (O/H)+12$<$7.2, $\log$ (N/O) at $\log$ (O/H)+12$=$6.5 is estimated to be -1.59$\pm$0.18. 

We perfome the Kolmogorov-Smirnov (KS) test, even though it is not likely to provide a meaningful statistical check because of the small number of samples. We determine the probabilities that the two data sets are drawn from the same population are less than 1\% and 9\% for each case of C/O and N/O, respectively. In summary, the statistical analysis including the random errors in the abundances,  suggests a relatively higher C/O and lower N/O ratios for DLAs than for EMP stars. However, this systematic difference remains uncertain until the potential systematic errors inherent in the stellar abundances are erased. Therefore, we await future work to improve the abundance determination with much more data before validating the argument presented in the following sections. Bearing in mind that the contrast between the relative abundances of C, N, and O claimed here must be viewed with caution, we attempt a theoretical interpretation of the stellar and DLA abundances.

\section{Analysis of C/O ratio}

As shown in the right panel of Fig. 1, the C/O ratio of stars clearly decreases as metallicity increases for $\log$ (O/H) + 12 $<$ 7.0. For theoretical interpretations of the origin of this feature, \citet{Chiappini_06} proposed a differential C production in accordance with the stellar rotation, while   \citet{Mattsson_10} proposed a time-dependent IMF. However, the former scenario predicts a more strongly decreasing trend of N/O than that of C/O because the stellar rotation influences N production more efficiently, which is not the case for the observed N/O trend (left panel). As for the latter case, there is no supporting evidence for it in other elemental ratios.  In addition, the claimed initially top-heavy IMF contradicts our present result, as we show in Sects. 5 and 6.

Here we present the theoretical interpretation of the C/O ratio observed in EMP stars, which gives a consistent view of the N/O trend and is developed into models of chemical evolution for the Galaxy and DLAs.

\subsection{Nucleosynthesis C/O}

We first see the nucleosynthesis results for C and O in massive stars, and show that the nucleosynthesis C/O ratio has a clear trend with respect to the progenitor mass. The left panels of Fig. 2 show the C and O yields in massive stars as a function of the progenitor mass, as calculated by three groups (circles, Woosley \& Weaver 1995; triangles, Nomoto et al.~1997; squares, Ekstr\"{o}m et al.~2008). There exist some differences in their models. The two former groups present the results of SN models, while the calculation by \citet{Ekstrom_08} does not include the stage of SN explosion. Since both C and O are not synthesized or broken by the SN shock, the three results can be equally compared. Another distinction is that the two SN models do not include the stellar rotation for the progenitor models, while the models by \citet{Ekstrom_08} include very fast rotation, which leads to a high C yield compared with the SN models. These comparisons reveal (i) good agreement with the mass-dependent O yield, but (ii) a variation in the C yield, partly due to  the difference in the stellar rotation.

Irrespective of the C variation in the different models, the resultant C/O ratios have a trend in common with the progenitor mass, i.e., a lower progenitor mass model results in a higher C/O ratio (right panel). Here, the C/O ratios denoted by [C/O] are normalized by a solar C/O ratio of -0.26. We emphasize that the range of the predicted [C/O] ratios with different progenitor masses covers an observed [C/O] range of -0.7 to 0 where there are EMP stars.

\subsection{Comparison with the observed C/O}

According to the claim based on the analysis of elemental abundance patterns of EMP stars \citep{McWilliam_95, Ryan_96, Audouze_95, Shigeyama_98, Nakamura_99, Umeda_02}, we assume that the formation of EMP stars is triggered by a SN remnant (SNR), thus the stars formed retain the abundance pattern of this SN. This hypothesis leads to the comparison of  the observed correlation of C/O and O/H for each EMP star with those of individual theoretical models. For this purpose, we calculate each model's [O/H] ratio obtained from the O yield in each SN model (lower left panel of Fig. 2) divided by the mass of hydrogen swept up by an SNR, as done by \citet{Shigeyama_98}. 

The mass $M_{\rm SW}$ swept up by an SNR is approximated by the formula \citep{Shigeyama_98}

\begin{eqnarray}
M_{\rm SW}=5.1 &\times& 10^4 M_\odot \left (\frac{E_0}{10^{51} \ {\rm ergs}}\right )^{0.97}  \nonumber \\
&\times& n_1^{-0.062}\left (\frac{C_s}{10 \ {\rm km \ s}^{-1}}\right )^{-9/7} \ \ \ , 
\end{eqnarray} 

\noindent where $E_0$, $n_1$, and $C_s$ denotes the explosion energy of SN, the number density of the ISM, and the speed of sound, respectively. In the models of \citet{Woosley_95}, $E_0$ varies from 1.28$\times 10^{51}$ to 3.01$\times 10^{51}$ ergs, while a single value of $10^{51}$ ergs is assigned to all models in \citet{Nomoto_97}. Since $M_{\rm SW}$ is insensitive to $n_1$, we determine the values of $M_{\rm SW}$ by adjusting $C_s$ so as to fit with the observed correlation  and our choice is  $C_s$=17 km s$^{-1}$ for the two SN models. On the other hand, for the models of massive stars by \citet{Ekstrom_08}, we adopt the values $E_0$= 10$^{51}$ ergs and $C_s$ = 6 km s$^{-1}$.

 \begin{figure}
   \centering
  \includegraphics[angle=0,width=0.45\textwidth]{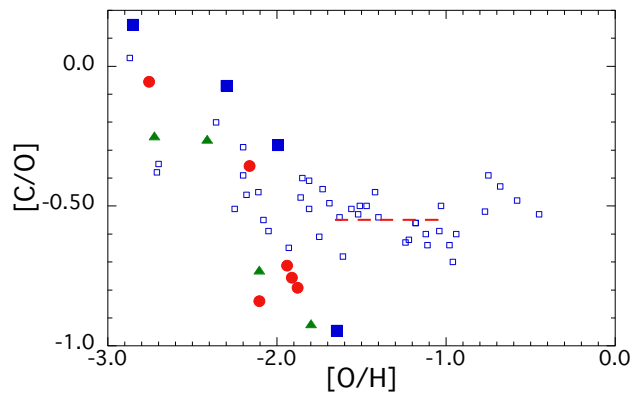}
       \caption{Comparison of the observed correlation of [C/O]-[O/H] for Galactic halo stars \citep[open squares:][]{Fabbian_09a} with the theoretical nucleosynthesis [C/O] ratio from massive stars as a function of the predicted [O/H]. The other symbols are the same as in Fig.~2. From each model,  SN models that reside in the range of -1$<$[C/O]$<+0.2$ and [O/H]$>$-3 are plotted. The [O/H] ratio is calculated from the ratio of the mass of O ejected from a SN to that of hydrogen swept by the SN remnant (see the text). For the theoretical model by \citet{Woosley_95}, the IMF-weighted [C/O] ratio of 12 \ms to 40 \ms SN models is denoted by the dashed line.      
}
  \label{}
   \end{figure}

The resultant C/O-O/H correlations predicted by individual SN models are compared with the observation in Fig. 3. From the results given by each model,  the points that reside in the range of -1$<$[C/O]$<+0.2$ and [O/H]$>$-3 are plotted. We see a good coincidence between the prediction and the observed data of EMP stars. Moreover, the average C/O ratio weighted by a Salpeter IMF utilizing the yields of \citet{Woosley_95} gives a good agreement with the C/O ratios for stars with -1.5$<$[O/H]$<$-1. Accordingly, we conclude that the decrease in C/O with increasing O/H is a reflection of the elemental feature of an individual SNR. In the end, the overall evolution of C/O imprinted in stellar abundances (i.e., squares) in Fig. 1 can be interpreted as follows. We can roughly divide its evolution into three stages that result in an increase in O/H: (i) for $\log$ (O/H)+12$<$7.0, the decreasing trend reflects the nucleosynthesis of the C/O ratio for individual SNe II; (ii) for 7.0$<$ $\log$ (O/H)+12$<$8.0, there is a plateau as a result of a mixture of C/O synthesized in various SNe II with different masses; and (iii) for $\log$ (O/H)+12$>$8.0, an increasing trend appears owing to a time-delayed additional C ejection from AGB stars.

The claim that we highlight is that a C/O ratio expected from individual SNe II is, theoretically and observationally, likely to be higher for a less massive progenitor mass. In this case, emergence of a kind of IMF, such as a cut-off of the upper mass end of massive stars results in a higher C/O ratio in the ISM at an early evolutionary stage, when only SNe II contribute to chemical enrichment. This  situation will be incorporated into the models of chemical evolution and shown to be a likely case for metal-poor DLAs.

\section{Chemical evolution model}

We first construct the model for Galactic chemical evolution (GCE) as a standard case. Subsequently, we try to reproduce the chemical evolution of other systems such as DLAs and extragalactic H II regions, modifying the GCE model by changing the fundamental model parameters, including the IMF and/or the star formation rate (SFR). 

\subsection{Galactic chemical evolution model}

The basic picture of the model is that the disk was formed through a continuous infall of material from outside the disk region over 14 Gyr. As is common in chemical evolution studies, we assume that the gas is distributed uniformly and that the heavy elements are well-mixed within the zone. In this framework, the SFR is assumed to be proportional to the gas fraction with a constant rate coefficient $\nu$ of 0.4 Gyr$^{-1}$ \citep{Tsujimoto_97, Tsujimoto_10}. For the infall rate, we adopt the formula that is proportional to $t\exp(-t/\tau_{\rm in})$ with a timescale of infall $\tau_{\rm in}$=5 Gyr \citep{Yoshii_96}. The metallicity of an infall is assumed to be [O/H]=-4 with a SN-II-like elemental ratio. In this study, we treat the evolution of C, N, and O. From the implication of the nucleosynthesis ratios in SNe II from EMP stars (see Fig.~1), we assume that $\log$ C/O =-0.6 and $\log$ N/O=-1.5 for infalling gas. Owing to a very small amount of metals in the infalling gas, both stellar and gas abundances for [O/H]$>$-3 (i.e., $\log$ (O/H)$+12>$-5.7) are unaffected by the abundances of an infall.

For the nucleosynthesis yields of SNe II, we adopt the results by \citet{Woosley_95} for C and O since their average nucleosynthesis C/O ratio is in good agreement with the observed C/O level. We discuss the N yield in Sect. 5. For the nucleosynthesis yields of type Ia SNe (SNe Ia), the O yield is  taken from \citet{Tsujimoto_95} with zero yields for C and N. In addition, we assume metallicity-dependent C and N yields from AGB stars. As a basic yield, we adopt the results calculated with $Z$=0.001, $n_{\rm AGB}$=4, and $m_{\rm HBB}$=0.8 by \citet{Hoek_97}, and we choose its metallicity dependence so as to follow the observed increasing feature seen in both C/O and N/O. These adopted nucleosynthesis yields are then combined with a power-law mass spectrum with a slope $x$ of -1.35, e.g., a Salpeter IMF, and the mass range (0.05 \msp, 40 \msp). The lower mass limit is deduced from the theoretical analysis of the mass-to-luminosity ratio in the solar neighborhood \citep{Tsujimoto_97}. For the occurrence frequency of SNe Ia, we assume that the fraction of the stars that eventually produce SNe Ia for $3-8$\ms is 0.05 with the lifetime of SN Ia progenitors of 0.5$-$3  Gyr \citep{Yoshii_96}. 

\subsection{Applications to other systems}

Our prime purpose is to reproduce the chemical evolution of DLAs. Based on the knowledge acquired by analysing the C/O ratio in Sect. 2, we introduce the IMF with the upper mass end at $<$ 40 \ms into the DLA case. In addition, we take into account the hypothesis that DLAs are metal-poor systems corresponding to dwarf galaxies \citep[e.g.,][]{Cen_03, Calura_03, Okoshi_05} or the outer disks of spiral galaxies \citep[e.g.,][]{Chen_05}. Accordingly, we assume a low SFR and a long timescale of infall for DLAs. 

  \begin{figure}
   \centering
  \includegraphics[angle=0,width=0.45\textwidth]{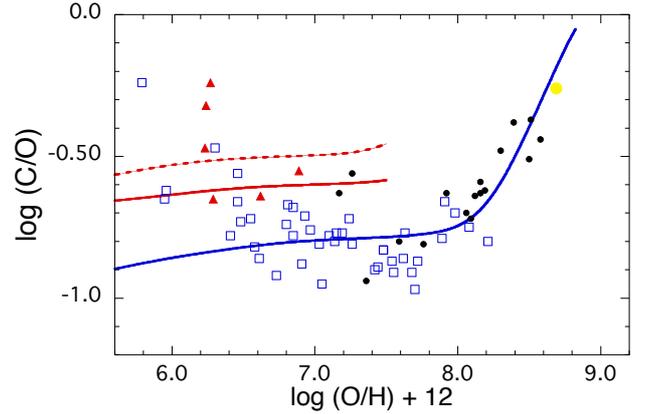}
       \caption{Predicted C/O evolution of the ISM against O/H with three different upper mass ends in the IMF, i.e., 40 \msp (blue solid curve), 25 \msp (red solid curve), and 20 \msp (red dotted curve). In the second and third models, the calculations stop when the metallicity reaches $\log$ (O/H)+12 = 7.5. The symbols of the observed data are the same as in Fig. 1.     
}
  \label{}
   \end{figure}

In the evolution of N/O, many extragalactic H II regions are likely to follow the path where chemical enrichment proceeds more efficiently than in the solar neighborhood and ends up with a more metal-rich ISM with $\log$(O/H)+12$>$9.0. In contrast to the DLA case, we assume a high SFR and a short timescale of infall for these metal-rich H II regions. The samples of HII regions with $\log$(O/H)+12$>$9.0 are indeed those in bright massive galaxies, mainly M31, M83, M51(an interacting galaxy), and NGC 4254, whose SFRs are expected to be higher than in the solar neighborhood. In addition, we introduce a top-heavy IMF, since the above choices do not agree sufficiently well with the observed trend of increasing N/O for these objects. In other words, since we find that the model with a high SFR and a normal IMF fails to reproduce the high N/O ratio observed in metal-rich H II regions, a top-heavy IMF is invoked (see Sect. 6).

In the end, we adopt IMFs that differ from Salpeter's for two objects to reduce or enhance the heavy-element yields ejected from SNe II. There are basically two approaches: varying the value of $m_u$ or changing the slope $x$. For the canonical case, $m_u$=40 \ms and $x$=-1.35 are assumed. Since a steeper IMF, i.e., $x<$-1.35,  does not give a satisfactory result because of its small effect on the elemental ratios for our adopted yields, $m_u$ is chosen as a variable parameter to reduce the yields. The restricted nucleosynthesis results lead to our choice of a flatter IMF, i.e., $x>$-1.35 to enhance the yields.

\section{C/O evolution}

The predicted C/O evolution for the Galaxy is shown by the blue solid curve in Fig. 4. It clearly reproduces the observational measurements for $\log$ (O/H) + 12 $>$ 7.0, which include those of  Galactic halo stars, extragalactic H II regions, and the Sun. The differences between the prediction and the observed data  seen for $\log$ (O/H) + 12 $<$ 7.0 occur because in such a  very metal-poor regime, stellar abundances (data of EMP stars) do not represent the abundance of the ISM (prediction) at each epoch when a star is formed, as already discussed. 

The observed data of DLAs can be directly compared with the model results. The C/O ratio in DLAs is expected to be higher than that of the ISM in an early Galaxy. To change the relative abundances  of C and O, we introduce an IMF with a variation in the upper mass limit, $m_u$. For two cases of $m_u$, 20\ms and 25\msp, the calculations are performed with the choice of parameters ($\nu$, $t_{\rm in}$)=(0.3 Gr$^{-1}$, 7 Gyr). The results are shown by the red dashed and solid curves, respectively. As expected from the mass-dependence of the nucleosynthesis C/O ratio shown in the right panel of Fig. 2, the smaller $m_u$ corresponds to a higher C/O ratio in the ISM that is compatible with the observed level of DLAs. Here we note that a smaller $m_u$ is found to lower the C/O ratio more efficiently than a steeper slope of the IMF, which is another mechanism for reducing the number of very massive stars.

\section{N/O evolution}

We continue our investigation of the N/O evolution, starting with the discussion of the N yield. In contrast to the case of the C yield, the nucleosynthesis results for N in massive stars vary significantly among different groups, thus we are far from converging on one reliable view. Therefore, we deduce the N yield as functions of the progenitor mass based on the theoretical argument about the observed N/O ratios of EMP stars, as well as  those of DLAs combined with the knowledge previously acquired from the discussion of the C/O ratio. This deduction has two major steps: 

(i) In contrast to the C/O feature, the N/O ratios of EMP stars apparently exhibit no clear trend. This  can be interpreted as an outcome of (1) very massive stars, the first contributor of chemical enrichment, making a broad level of the observed N/O ratio, and (2) their prior contamination of the ISM with a high N/O ratio that prevents a lower N/O ratio from less massive stars from being established in stellar abundances. In other words, the N/O ratios of EMP stars roughly correspond to  the nucleosynthesis N/O in very massive stars while the information on less massive stars is not established because of  with a time delay their small amount of N production relative to more massive stars. Recent nucleosynthesis results \citep{Hirschi_07, Ekstrom_08} support the idea that the N/O ratios in very massive stars broadly give the observed level of the N/O ratio for EMP stars. 

(ii) After detecting the low N production in less massive stars and fixing the yield at the most massive mass, the small amount of N synthesized in less massive stars is not so critical. We can estimate  the amount by referring to nucleosynthesis results and with supplementary information about the discussion of the C/O ratio: the abundances of DLAs are likely to reflect the nucleosynthesis products from less massive ($<$ $\sim$20-25\msp) stars. Thus, we assume that  the nucleosynthesis products in 20-25 \ms imply that log (N/O)$\sim$ -2.

\noindent Through these assessments, we finally obtain the N yield shown in the upper panel of Fig. 5.

  \begin{figure}
   \centering
  \includegraphics[angle=0,width=0.45\textwidth]{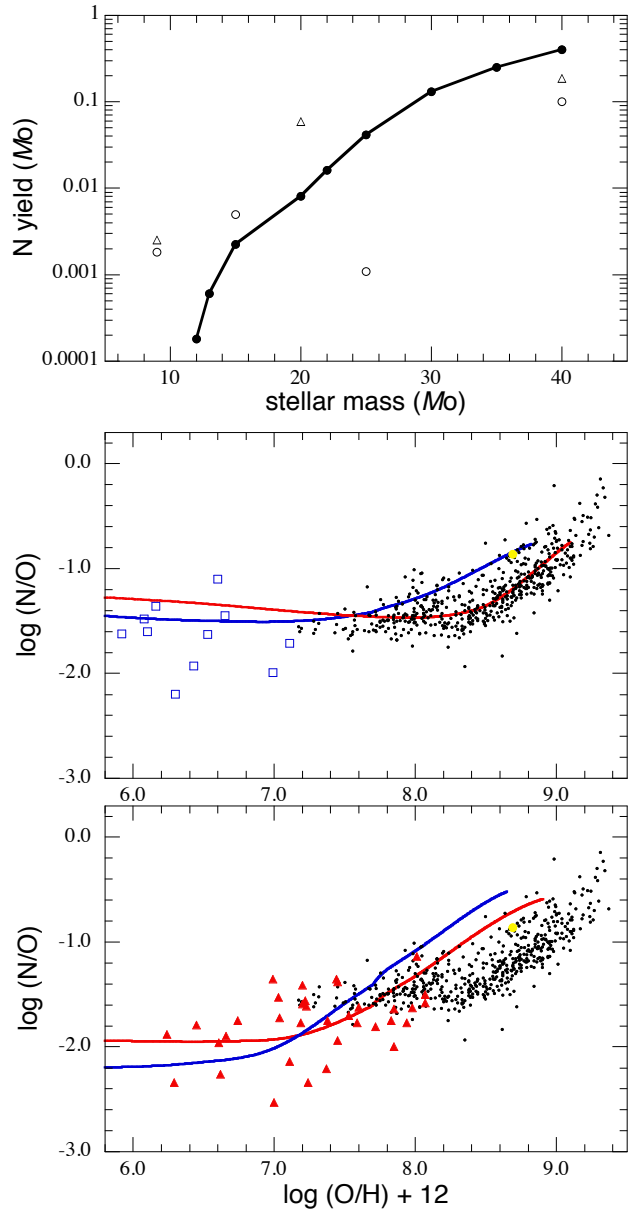}
       \caption{{\it upper panel}: Adopted N yield as a function of the progenitor mass. For reference, new nucleosynthesis results are attached (circles: Ekstr\"{o}m et al. 2008; triangles: Hirschi 2007). {\it middle panel}: N/O evolution predicted by the models with a Salpeter IMF ($x$=-1.35; blue curve)  and a flatter one ($x$=-1.1; red curve). The symbols of the observed data are the same as in Figure 1. {\it lower panel}: N/O evolution calculated with the models in which the upper mass of the IMF ends at 25 \ms (red curve) and 20 \ms (blue curve). 
}
  \label{}
   \end{figure}

Utilizing the deduced N yield, we first model the metal-rich systems aiming for the Galaxy and extragalactic H II regions, and show their results by blue and red curves, respectively, in the middle panel of Fig. 5. For the extragalactic H II regions, we set  ($\nu$, $t_{\rm in}$)=(5 Gyr$^{-1}$, 1 Gyr), together with a flatter IMF of $x$=1.05. The choice of a top-heavy IMF, in addition to a high rate of star formation, is necessary to accelerate chemical enrichment during an early phase: the N ejected from AGB stars starts to occur on a very short timescale ($\sim 4\times 10^7$yr) because the major source of N,  AGB stars, are massive (6-8 \msp). As an effect of a top-heavy IMF, the N/O ratio in an early phase becomes higher than the Galaxy case. 

We emphasize that HII regions with a relatively low $\log$(N/O) for $\log$(O/H)+12\gtsim 8.0 are reproduced by the model with a flatter IMF. We know from observations that the more luminous disk galaxies are more metal-rich because they are redder \citep[e.g.,][]{Grijs_99}. It is thus possible to claim that more luminous disk galaxies are likely to have a flatter IMF. Here, we propose that the N/O ratio in individual disk galaxies at a given metallicity is the more stringent parameter needed to deduce their IMFs. Investigation of the correlation between the N/O ratio and physical properties of disk galaxies such as the mean mass density and the total mass will surely provide a clearer understanding of the determinant factor for the IMF in disk galaxies. In addition, HII regions with a flatter IMF are predicted to exhibit a lower C/O ratio on the whole than those with a canonical IMF for the same reason, which will be confirmed by future observations.

The opposite situation corresponding to metal-poor DLAs is considered in the lower panel of Fig. 5. 
The N/O evolution is calculated with the models in which the upper mass in the IMF ends at 25 \ms (red curve) and 20 \ms  (blue curve). Both models closely reproduce the overall chemical evolution of DLAs. Their most conspicuous property is represented by an early and steep increase in N/O owing to the ejection of N from AGB stars. This property is caused by the suppression of the ejection of O from SNe II because the IMF lacks very massive stars, while the contribution of N  from AGB stars remains at  the same level. In addition, we see a large scatter in the N/O ratio of DLAs, some of which deviate from our predicted paths. This may reflect the diverse chemical evolution of individual DLAs--- some  DLAs may have a canonical IMF, while some chemical enrichment proceeds with a higher SFR. 

We add one comment. The low N abundance in DLAs is reminiscent of the Magellanic Clouds (MCs), which exhibit a severe deficiency of N in H II regions \citep[e.g.,][]{Hill_04}. However, an IMF without very massive stars, as predicted in metal-poor DLAs, is not likely to be representative of the MCs.  Our prediction is that the upper-mass cut-off causes a low N/O but a high C/O, whereas the MCs also exhibit a deficiency of C abundance. Moreover, the low N/O occurs only in an early phase, before AGB stars begin to contribute to the N enrichment. Neither is a top-heavy IMF, which accelerates chemical enrichment and thereby results in the low N/O for young H II regions. \citet{Bekki_10} propose that the origin of a low N/O in the MCs may be caused by an external factor, i.e., a dilution by  a gas with a severely deficient N abundance such as high-velocity clouds, an idea that needs more research.

\section{Discussion}

\subsection{Other evidence of a non-universal IMF}

Although we first propose that the formation of very massive stars (masses higher than $\sim$ 20 \msp) is truncated in DLAs,  observations have suggested that the formation of massive stars is truncated or significantly suppressed by the IMF with a steeper slope in some local galaxies. \citet{Meurer_09} investigated the extinction-corrected flux ratio $F_{\rm H \alpha}/f_{\rm FUV}$ (where $F_{\rm H \alpha}$ and $f_{\rm FUV}$ are the observed H$\alpha$ flux and far-ultraviolet one, respectively) in various galaxies and found that the ratio is smaller for galaxies with lower surface-brightnesses in the $R$-band (see their Fig. 6). They considered the possibility that $F_{\rm H \alpha}$ comes mostly from massive O-type stars with masses higher than 20 \ms that can ionize the ISM, while $f_{\rm FUV}$ comes from less massive stars. In the end, they concluded that the observed correlation results from there being a smaller number  of massive O-type stars in lower surface-brightness galaxies. That led to their claim of a nonuniform IMF for galaxies in the local Universe. Furthermore, \citet{Hoversten_08} investigated the IMFs for a large number of galaxies with different magnitudes in a database of the Sloan Digital Sky Survey and found that the IMFs in bright galaxies such as the Milky Way are similar to the Salpeter IMF, whereas fainter galaxies tend to have steeper IMFs. Their results imply that fewer massive stars are formed in low-luminosity galaxies. 

The possible truncation of the formation of massive stars in low-luminosity galaxies and LSB galaxies is discussed for other samples of galaxies \citep[e.g.,][]{ Boselli_09, Lee_09, Lee_11b}. \citet{Boselli_09} claimed that the flux ratio $F_{\rm H \alpha}/f_{\rm FUV}$ is barely related to galaxy properties such as stellar mass and effective surface brightness in late-type galaxies, and argued against the truncation of stars in these galaxies. In addition, the results by \citet{Lee_11b} implied that a deficiency of H$\alpha$ possibly results from Poisson fluctuations in the formation of massive stars for a universal IMF. Thus, it is safe to say that it would be premature to claim that the truncated IMF is characteristic of low-luminosity galaxies and LSB galaxies. 

From a theoretical aspect, the high-mass end ($m_{\rm u}$) has been found to depend on the mass of star clusters, thus on the physical properties of galaxies where the clusters are formed \citep[e.g.,][]{Pflamm_08}. If more massive clusters  are less efficiently formed in the ISM with a lower gas density as in the outer parts of galaxies, $m_{\rm u}$ would be lower \citep{Pflamm_08}. This theoretical result appears to support a scenario where $m_{\rm u}$ is inclined to be lower in less massive and LSB galaxies of low mean mass-density and thereby the formation of massive star clusters is suppressed.

To explain the observed physical properties of massive galaxies,  previous studies, both observational and theoretical, pointed out that the IMFs need to be top heavy \citep[e.g.,][]{Nagashima_05, Loewenstein_06, Dokkum_08}. For example, the observed $\alpha$/Fe in typical $L_{\star}$ ellipticals can be reproduced by semi-analytic models based on the hierarchical galaxy formation scenario only if  top-heavy IMFs are adopted \citep[][who, however, find a poor fit to the Mg/Fe - $\sigma$ relation]{Nagashima_05}. The ISM of these massive galaxies with top-heavy IMFs at high redshifts are  predicted to exhibit low C/O and high N/O ratios, in contrast to the properties of high-$z$ DLAs. In this respect, massive galaxies are unlikely to be identified with DLAs.

Several theoretical works have suggested that the slope of IMFs in galaxies evolves with time in the  sense that the slope becomes flatter, i.e., top-heavier in accordance with the growth of the galaxy \citep[e.g.,][]{Larson_98}. If we take into account the dependence of the IMFs on the physical properties of galaxies, we find that the formation of massive stars is correlated with the gas mass density of the ISM. We are then led to a picture where, as galaxy masses grow and surface mass densities become higher, galactic IMFs change from  truncated IMFs into canonical ones. We recall, however, that the chemical abundances of the local halo and disk stars show no evidence of a variable IMF in the past. 

Finally, there is strong evidence of a top-heavy IMF in our own Galaxy. The chemical properties of the Galactic bulge stars, especially their stellar metallicity distribution, are most consistent with a top-heavy IMF \citep[e.g.,][]{Matteucci_90, Ballero_07, Tsujimoto_10}. 

\subsection{Origin of  the possible truncated IMF in DLAs}

As discussed  above,  $m_{\rm u}$ can be smaller in dwarf galaxies and LSB galaxies.  Our prediction is that a small $m_{\rm u}$ results in the relatively high C/O and low N/O ratios in the ISM. Therefore, this result is expected in these galaxies at an early epoch of their evolution. After all, the observed abundance pattern for high-z DLAs \citep[][]{Pettini_08} suggests that DLAs are gas-rich dwarf galaxies or LSBs. In addition, the gas-rich outer parts of high-$z$ disk galaxies represent another possible explanation of DLAs since the local outer disks may have lower $m_{\rm u}$ and thus lack H$\alpha$ regions \citep{Pflamm_08}. Future observations designed to measure the N/O and C/O ratios in the outer HII regions of high-$z$ disk galaxies will validate their connection with DLAs, as well as their potential to have truncated IMFs. we note that owing to a low $m_{\rm u}$ in gas-rich dwarfs and the outer part of disk galaxies, the formation of HII regions is significantly suppressed so that their HI-rich ISM can be identified with DLAs. The small $m_{\rm u}$ would increase the possibility of galactic  ISM being detected as HI-rich DLAs. 

For UV-selected galaxies, \citet{Contini_02} reveal the correlation of the N/O ratio with the luminosity. In their samples,  galaxies exhibiting log N/O $<$ -1.6 are restricted to fainter ones with $M_{\rm B}>$-18, which correspond to dwarf galaxies. According to our present study, the correlation is caused by the IMF lacking massive stars in dwarf galaxies. In addition, the similarity of the N/O ratio in dwarf galaxies and DLAs supports the idea that DLAs are gas-rich parts of dwarf galaxies at high redshifts. DLAs can indeed be identified with dwarf galaxies \citep[e.g.,][]{Cen_03, Calura_03, Okoshi_05}.
  
\section{Conclusions}

We have presented a new view of the chemical evolution of DLAs as well as extragalactic H II regions  from our careful analysis of their C, N, and O abundances. We highlight two nucleosynthesis results: (i) the C/O ratio has a clear correlation with the progenitor mass in the sense that a less massive SN II yields a higher C/O, and (ii) the finding of a new channel for the very high N production in the most massive stars. These combined with the hypothesis that the abundances of EMP stars in the Galactic halo reflect those of individual SN remnants naturally lead to a theoretical interpretation for the abundances of DLAs. Metal-poor DLAs have puzzling properties, a relatively high C/O ratio and relatively low N/O ratio compared to EMP stars. Our claim is that these properties are compelling evidence that the IMF lacks very massive stars in metal-poor DLAs. 

In contrast,  we have found that some metal-rich extragalactic H II regions in local disk galaxies appear to have top-heavy IMFs. This suggests that the IMFs are likely to differ even among disk galaxies. There were earlier claims that high-$z$ early-type galaxies possess a top-heavy IMF \citep[e.g.,][]{Dokkum_08}. Our additional findings resulting from a detailed chemical analysis focusing on DLAs strengthen supporting evidence of a non-universal IMF from high to low redshifts in the Universe. However, on the other hand, there is extensive evidence of a universal IMF \citep{Bastian_10}. Future work will definitely be required to validate the IMF variation. Finally, we  emphasize that analysing the C/O and N/O abundances more precisely in galaxies, including our own, will help us to understand the IMF variation.

\acknowledgements
The authors wish to thank an anonymous referee for all valuable comments that helped improve the paper. TT is assisted in part by Grant-in-Aid for Scientific Research (21540246) of the Japanese Ministry of Education, Culture, Sports, Science and Technology.

\end{document}